\documentclass[journal, 12]{IEEEtran}
\usepackage[pdftex]{graphicx}
\usepackage[cmex10]{amsmath}
\usepackage{amsfonts}
\usepackage[english]{babel}
\usepackage{amssymb}
\newcommand{\q}[1]{{\bf #1}}
\usepackage{cite}
\allowdisplaybreaks
\usepackage{algorithm}
\usepackage{algorithmic}
\usepackage[font={small}]{caption}
\newcommand{\snir}{{{\Gamma}}}
\newcommand{\snr}{{{\Gamma}}}

\begin{document}

\title{IRS-Enabled Backscattering in a Downlink Non-Orthogonal Multiple Access  System}

\author{  Azar Hakimi, Shayan Zargari, Chintha Tellambura, \IEEEmembership{Fellow, IEEE,}   Sanjeewa Herath, \IEEEmembership{Member, IEEE}
\thanks{A. Hakimi, S. Zargari, and C. Tellambura with the Department of Electrical and Computer Engineering, University of Alberta, Edmonton, AB, T6G 1H9, Canada (e-mail: {hakimina, zargari, ct4}@ualberta.ca).  \\
\indent S. Herath is with the Huawei Canada, 303 Terry Fox Drive, Suite 400, Ottawa, Ontario K2K 3J1 (e-mail: sanjeewa.herath@huawei.com).}  }
 
\maketitle

%%%%%%%%%%%%%%%%%%%%%%%%%%%%%%%%%%%%%%%%%%%%%%
                %ABSTRACT%
%%%%%%%%%%%%%%%%%%%%%%%%%%%%%%%%%%%%%%%%%%%%%%              
\begin{abstract}
Intelligent reflecting surface (IRS)-enabled backscatter communications can be enabled by an access point (AP) that splits its transmit signal into modulated and unmodulated parts. This letter integrates non-orthogonal multiple access  (NOMA) with this method to create a two-user primary system and a secondary system of IRS data.  Considering the decoding order, we maximize the rate of the strongest primary user by jointly optimizing the IRS phase shifts, power splitting (PS) factor at the AP, and NOMA power coefficients while guaranteeing the quality of service (QoS) for both weak user and IRS data in the primary and secondary systems, respectively. The resulting optimization problem is non-convex. Thus, we split it into three parts and develop an alternating optimization (AO) algorithm. The advantage is that we derive closed-form solutions for the PS factor and NOMA power coefficients in the first two parts. In the third part, we optimize the phase shifts by exploiting semi-definite relaxation (SDR) and penalty techniques to handle the unit-modulus constraints. This algorithm achieves substantial gains (e.g., 40--68\%)  compared to relevant baseline schemes.
\end{abstract}


%%%%%%%%%%%%%%%%%%%%%%%%%%%%%%%%%%%%%%%%%%%%%%
                %INTRODUCTION%
%%%%%%%%%%%%%%%%%%%%%%%%%%%%%%%%%%%%%%%%%%%%%% 
\section{Introduction}
We consider the problem of simultaneously transmitting data through an intelligent reflecting surface (IRS) and the data from an access point (AP) to two users (Figure 1).  The  IRS not only backscatters its own data (secondary data) that can come from embedded sensors and similar devices but also passively reflects the AP signals (primary data) \cite{Park2020, zhao2020}. This capability is helpful in smart homes, smart cities, and  Internet of Things (IoT) networks,  where heterogeneous sensors \cite{XuXWHY22},  network technologies, and communication protocols improve energy efficiency, spectral efficiency, and other metrics.  Although ambient backscatter communication (AmBC) may achieve the same task,  strong direct-link interference greatly impacts the backscatter signal detection \cite{Rezaei2020}.

How does the IRS modulate its data onto the AP signal and ensure that the IRS-data signal does not interfere with the decoding processes at $U_1$ and $U_2$?  To address these issues, we borrow a concept developed in \cite{Park2020,GuoLL19a}. Specifically, \cite{Park2020} shows that if the AP  simultaneously transmits an unmodulated carrier and modulated carrier, then the IRS can backscatter its own data and reflect the AP signals too. Thus, the AP splits its transmit power between these two carriers according to the power splitting (PS)  ratio $\alpha $ (where $0 \leq \alpha \leq 1)$. The modulated carrier transports the AP data for $U_1$ and $U_2$, and the unmodulated carrier is modulated by the IRS to send its own data, which is broadcast to both users.  Since the rate of IRS data is much lower than that of primary system, simple filtering can help $U_1$ and $U_2$ to decode the IRS data first and subtract it to avoid interference in decoding data. The process also supports data transmissions without requiring perfect symbol synchronization\cite{GuoLL19a}, which is important for low-complexity and low-cost IoT devices.

When $U_1$ is decoding its data, the data of $U_2$ and that of the IRS act as interference. Thus, we leverage non-orthogonal multiple access (NOMA) to differentiate between $U_1$ and $U_2$. Specifically, NOMA can do that by exploiting the channel gain disparities (see \cite{YangXLR21} and references therein) and successive interference cancellation (SIC) decoding methods. IRS and NOMA together can boost the capacity, coverage, spectral efficiency,  and other quality of service (QoS) metrics  \cite{mu2019exploiting}.

Previously,  \cite{GuoLL19a} was the first paper that proposed splitting the AP signal into unmodulated and modulated carrier signals.    Using this modified AP signal,  \cite{Park2020} developed the concept of the IRS-enabled backscattering.  However, these works \cite{Park2020,GuoLL19a} differ from our study in several ways. The focus of  \cite{GuoLL19a}  is to develop a cognitive backscatter system and an associated cooperative receiver. Thus, the user can decode both the AP data and backscattered data successfully.  Neither an IRS nor NOMA is considered in this paper. It is limited to a single backscatter device. The focus of \cite{Park2020}  is to enlist an IRS to send additional secondary data.  The conceptual novelty of our paper compared to \cite{Park2020,GuoLL19a} is the integration of downlink NOMA  with  IRS-enabled backscattering.

Our work also differs from existing works \cite{Jia2020, WZhao2020}, where the IRS assists in backscattering the AP signals. Further, our work differs from \cite{XuQGGSA21} where two NOMA users can receive a version of the BS signal reflected from a backscatter device. Also, in \cite{YangXLR21}, authors use the IRS to assist the downlink NOMA users. In this letter, we fully optimize the system depicted in Figure 1. To this end, we jointly optimize the PS factor, NOMA power coefficients, and IRS phase shifts to maximize the rate of the stronger user. The problem is formulated by considering the decoding order of the users and guaranteeing QoS for both primary and secondary systems. However, the problem is non-convex,  and widely available convex optimization techniques do not help.    

We thus utilize the alternating optimization (AO) method. It is simply an iterative procedure for maximizing a function jointly over all variables by
alternatively maximizing over individual subsets of them.  Therefore, we split the variables into three subsets, namely the PS factor, NOMA power coefficients, and IRS phase shifts. We derive closed-form solutions for the PS factor and NOMA power coefficients for the first and second subproblems, respectively. For the phase shifts, we use semi-definite relaxation (SDR) and penalty techniques to optimize them subject to their unit-modulus constraints. We combine these three tasks into an overall algorithm (\textbf{Algorithm \ref{algorithm_AO}}).  We compare it to several baseline schemes, including one with orthogonal multiple access (OMA). The performance gains are 40-68\%. 

\textit{Notation:} Vectors and matrices are indicated by boldface lowercase and capital letters. For a square matrix $\mathbf{A}$, $\mathbf{A}^H$, $\mathbf{A}^T$, $\text{Tr}(\mathbf{A})$, $||\mathbf{A}||_{*}$, and $\text{Rank}(\mathbf{A})$ denote its Hermitian conjugate transpose, transpose, trace, trace norm, and Rank, respectively. $\mathbf{A}\succeq\mathbf{0}$ denotes a positive semidefinite matrix. $\text{diag}(\cdot)$ is the diagonalization operation. The Euclidean norm  and the absolute value are  $\|\mathbf x \|$ and $|x |$. $\nabla_{\mathbf{x}}f(\mathbf{x})$ is the  gradient vector over  $\mathbf{x}$. The expectation operator is $\mathbb{E}[\cdot]$. A circularly symmetric complex Gaussian (CSCG) random vector with mean $\boldsymbol{\mu}$ and covariance matrix $\mathbf{C}$ is denoted by $\sim \mathcal{C}\mathcal{M}(\boldsymbol{\mu},\,\mathbf{C})$. Besides, $\mathbb{C}^{M\times N}$ indicates $M\times N$ dimensional complex matrices.

%%%%%%%%%%%%%%%%%%%%%%%%%%%%%%%%%%%%%%%%%%%%%%
                %SYSTEM MODEL%
%%%%%%%%%%%%%%%%%%%%%%%%%%%%%%%%%%%%%%%%%%%%%% 
\section{System Model}\label{system}

\begin{figure}[t]
    \centering
   \includegraphics[width=3in]{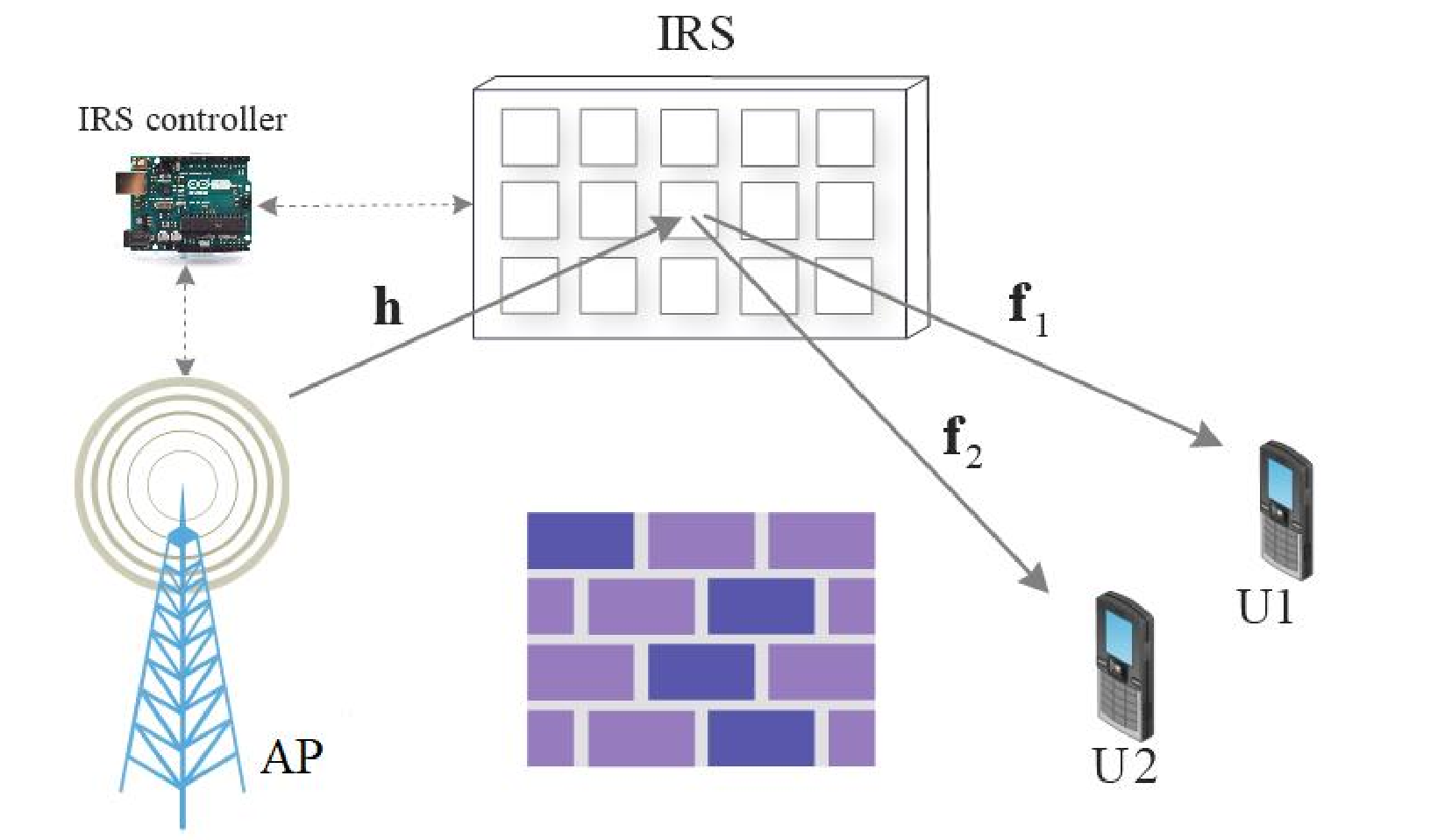}
    \caption{IRS-aided NOMA AmBS system model.}
    \label{SM}   
\end{figure}

Figure \ref{SM} depicts the considered  IRS-aided NOMA AmBC consists of a single-antenna AP operating based on the NOMA scheme, an $M$-passive reflecting elements IRS indexed by  $m\in\mathcal{M}=\{1,\dots, M\}$, and two single-antenna users in the set of $\mathcal{K}=\{1,2\}$. We assume that there is no direct link between the  AP and users due to high attenuation and signal blockages, a typical scenario \cite{Jiakuo_Zuo,WZhao2020,FangXPD20}. This assumption holds for urban environments and  above-6~GHz frequency bands such as millimeter wave, where signal blockages are frequent. 
In Figure~\ref{SM}, the   IRS is also backscattering its sensed data.  As mentioned before,  the AP transmits modulated and unmodulated carriers simultaneously and divides its power between these two parts.  Therefore,  the AP  baseband transmit signal  at time instant $n$ can be represented as 
\begin{equation}
x(n) = \sqrt{\left(1-\alpha\right)P_\text{T}}+\sqrt{\alpha P_\text{T}} \sum_{k=1}^{2}a_{k} s_k(n), 
\end{equation}
where $\alpha$ is the PS factor for the AP, $P_\text{T}$ denotes the AP transmit power, $s_k$ is the $k$-th user data that  satisfies $\mathbb{E}\left[|s_{k}(n)|^2\right]=1,~k\in\mathcal{K}$, and $a_k$ is the  NOMA power allocation coefficient of the $k$-th user.  To ensure fairness, the user with lower channel gain gets assigned a    higher $a_k.$

The spectrum of the AP RF  signal consists of a wideband message signal and an unmodulated carrier signal \cite{Park2020}.  Upon receiving the RF  signal, the IRS can modulate the unmodulated part, e.g.,  $\sqrt{(1-\alpha)P_T} e^{jw_ct}$,  to send its own data using binary phase-shift keying (BPSK) modulation \cite{Park2020}.

Denote channels from the AP-to-IRS and IRS-to-user $k$ as $\q h\in \mathcal{C}^{1\times M}$ and $\q f_{k}\in\mathcal{C}^{M\times 1}$, $\forall k\in \mathcal{K}$, respectively. All the channels undergo quasi-static flat Rician fading and remain unchanged for several symbols  \cite{Shayan_zar3}. We also assume the channel state information (CSI) availability. These are standard assumptions in the literature. This system setup has the added advantage of easy synchronization because the time delay can be compensated through the passive reflecting elements at the IRS \cite{Park2020,zhao2020}.  Let $\mathbf{\Theta}= \textrm{diag}\left(\beta_{1}e^{j\theta_1},\dots, \beta_{m}e^{j\theta_m},\dots,\beta_{M}e^{j\theta_M}\right)$ represents the reflection coefficient matrix at the IRS, where $\beta_{m}\in[0,1]$ and $\theta_m\in[0,2\pi)$,  $\forall m\in\mathcal{M}$, are the reflection amplitude and phase shift of the $m$-th passive reflecting element at the IRS, respectively. To maximize the reflected signal power at the IRS and also to mitigate the hardware cost, we assume that $\beta_{m}=1, \forall m\in\mathcal{M}$. Consequently, the received signal at each user can be written as 
\begin{equation}
y_k(n) = \q h\mathbf{\Theta}\q f_k x(n) +z_k(n),~\forall k\in \mathcal{K},
\end{equation}
where $z_k\sim \mathcal{C}\mathcal{M}(0,\sigma^2)$, $\forall k\in \mathcal{K}$, is the received noise with variance $\sigma^2$, and it is assumed to be the same for all users.
 
\section{Transmission Scheme}
This works as follows. First, the AP  transmits $x(n)$ with the symbol period $T_s$. Second, the IRS manipulates the unmodulated part of $x(n)$  to transmit its data, $b(n)$, by applying BPSK modulation with bit period $T_b\gg T_s$. For $b(n) = 1$ or $0$, the IRS adds the following phase shifts: $0$ or $\pi$.  As a result, each user see  the original  signal when $b(n) = 1$, and a negative of it   when $b(n) = 0$. Since the primary system rate is much higher than the IRS data rate  ($T_b = LT_s$, where  $L \gg 1$), each user can decode the IRS data with a simple band-pass filter as each  IRS data bit remains constant over $LT_s$ duration.   Accordingly, each user first decodes the IRS data, cancels the effect of the  IRS data, and decodes its own data. The IRS data symbol is $L$ times longer than the nominal user data symbol.   Because each user must decode the IRS data symbol first before decoding its own data, this process will introduce a delay of $L$ bits.

Thus, the received signal-to-noise ratio (SNR) of the secondary signal at each user is given by
\begin{align}
    \snr_{c_k} = \frac{L(1-\alpha)P_\text{T}|\q h\mathbf{\Theta}\q f_{k}|^2}{\sigma^2} ,~\forall k\in \mathcal{K}.
\end{align}
Consider $\Phi(k) \in\{0,1\}$ as the decoding order of user $k$, where $\Phi(k) = 1$ and $\Phi(\bar{k})=0$, $\forall k,\:\bar{k}\in \mathcal{K}$ indicate that the signal of user $\bar{k}$ is first decoded by treating user $k$’s signal as interference. Then, by removing user $\bar{k}$’s signal via SIC, user $k$ decodes its signal without the co-channel interference. Explicitly, it means that $|\q h \mathbf{\Theta} \q f_k |^2\geq |\q h \mathbf{\Theta} \q f_{\bar{k}}|^2$. On the other hand, for $\Phi(k) = 0$ and $\Phi(\bar{k})=1 $, we have $|\q h \mathbf{\Theta} \q f_{\bar{k}} |^2\geq |\q h \mathbf{\Theta} \q f_{k}|^2$. Accordingly, the required signal-to-interference-plus-noise ratio (SINR) of user $k$ can then be represented as
\begin{align}\label{SINR}
    \snir_{k} = \frac{\alpha P_\text{T} a_k|\q h\mathbf{\Theta}\q f_{k}|^2}{ \Phi(\bar{k})\alpha P_\text{T} a_{\bar{k}}|\q h\mathbf{\Theta}\q f_{{k}}|^2+\sigma^2} ,~\forall k, \bar{k}\in \mathcal{K}.
\end{align}
Subsequently, the  rate achieved  at user $k\in \mathcal{K}$ based on the Shannon capacity can be expressed  as $R_{k} = \log(1+\snir_{k})$.
 
%%%%%%%%%%%%%%%%%%%%%%%%%%%%%%%%%%%%%%%%%%%%%%
                %Problem Formulation%
%%%%%%%%%%%%%%%%%%%%%%%%%%%%%%%%%%%%%%%%%%%%%% 
\section{Problem Formulation}\label{problem:formulation}

We maximize the rate of the strong user while providing the QoS required for the weak user to decode its data. To this end, we jointly design $\mathbf{\Theta}$, $\alpha$, and $a_k$, $k\in \mathcal{K}$, while considering the decoding order. Let $\Phi(k)$ be an indicator function of the stronger user (i.e., one with a larger channel gain).  That is, if   $\Phi({k}) = 1$ and $\Phi(\bar{k}) = 0$, then  user $k$ is the stronger user. We formulate the  optimization problem as follows:
\begin{subequations}
\begin{align}\label{opt:prob1}
 \hspace{-3mm} (\mathrm{P1}):  &\max_{ \mathbf{\Theta},a_k,\alpha} \hspace{0.5em} R_k = \log_2(1+\Gamma_k), \\
 &  \text{s.t.}  \quad
      |\q h \mathbf{\Theta} \q f_k|^2\geq |\q h \mathbf{\Theta} \q f_{\bar{k}}|^2,~\text{for}\hspace{0.5em} \Phi(k)=1, \Phi(\bar{k}) = 0,\label{opt1:1}\\
   &\quad\quad  \snir_{\bar{k} }\geq \gamma_{\text{th}} ~\text{for}\hspace{0.5em} \Phi(k)=1,  \label{opt1:2}\\
   &\quad \quad \snr_{c_k} \geq \gamma_{\text{th}} ,~\forall k\in\mathcal{K},\label{opt1:33} \\
   &\quad  \quad |e^{j\theta_m}|=1 ,~\forall m\in\mathcal{M}, \label{opt1:4}\\
   &\quad  \quad 0 \leq \alpha\leq 1, \label{opt1:5}\\
   &\quad  \quad a_1+a_2=1, \label{opt1:6}
\end{align}
\end{subequations}where (\ref{opt1:1}) indicates the decoding order at the NOMA users, and (\ref{opt1:2}) as well as (\ref{opt1:33}) are the QoS constraints that indicate the minimum SINR requirement of the weak user of the primary system and secondary system, respectively. (\ref{opt1:4}) denotes the unit-modulus constraints of the phase shifts at the IRS. Finally, (\ref{opt1:5}) and (\ref{opt1:6})  are the natural  limits  for $\alpha$ and $a_k.$ 

We hasten to add that other optimization criteria include the weighted sum rate, max-min rate, and others \cite{Jiakuo_Zuo, mu2019exploiting}.  These may be developed in future works.

\section{Proposed Solution}\label{proposed}
To solve $(\mathrm{P1})$ efficiently, we employ the AO method. AO is a widely used approach to attack non-convex problems  \cite{mu2019exploiting, FangXPD20, Shayan_zar1,Shayan_zar3}. The AO algorithm optimizes one block of variables at a time while keeping other blocks fixed. Following this approach, we break the problem into three simpler sub-problems. We derive closed-form expressions for $ \alpha$  and $a_k,~\forall k\in \mathcal{K}$ in the first and second subproblems, respectively. For $\mathbf{\Theta}$, we use SDR  and penalty techniques by invoking successive convex approximation (SCA) to optimize them.
Although the Gaussian randomization can provide rank one candidates, they may be infeasible for the original problem and may impose performance losses. That is why we choose a  penalty factor to ensure the rank-one constraint. 

Since the objective function in ($\mathrm{P1}$) is a  monotonically increasing $\log(\cdot)$ function and $\Gamma_k$ for $\Phi(k) = 1$ inside the $\log(\cdot)$ is a linear function of $\alpha$, we conclude that the objective is an increasing function over  $\alpha$. Thus, $\alpha$ achieves its optimum value in a  corner of its feasible regime.  Based on \eqref{opt1:33}, the optimal value of $\alpha$ can be obtained as 
\begin{align}\label{alpha_opt}
    \alpha_{\text{opt}} = \min\left(1-\frac{\sigma^2\gamma_{\text{th}}}{LP_\text{T}|\q h \mathbf{\Theta} \q f_1|^2},~1-\frac{\sigma^2\gamma_{\text{th}}}{LP_\text{T}|\q h \mathbf{\Theta} \q f_2|^2}\right).
\end{align}
However, to have a feasible regime for $\alpha$, the following conditions also need to be satisfied:
\begin{align}
     &\frac{\sigma^2 \gamma_{\text{th}}}{a_2 P_\text{T} |\q h \mathbf{\Theta} \q f_2|^2\!-\! a_1 P_\text{T} \gamma_{\text{th}} |\q h \mathbf{\Theta} \q f_2|^2} \leq\! \alpha_{\text{opt}},\\ 
      &\frac{\sigma^2 \gamma_{\text{th}}}{a_1 P_\text{T} |\q h \mathbf{\Theta} \q f_1|^2\!- \!a_2 P_\text{T} \gamma_{\text{th}} |\q h \mathbf{\Theta} \q f_1|^2}\leq\! \alpha_{\text{opt}},
\end{align}
if $|\q h \mathbf{\Theta} \q f_1|^2\geq |\q h \mathbf{\Theta} \q f_{2}|^2$ and $|\q h \mathbf{\Theta} \q f_2|^2\geq |\q h \mathbf{\Theta} \q f_{1}|^2$, respectively. Consequently, as a closed-form solution is derived for $\alpha$, one can remove it from the optimization variables in $(\mathrm{P1})$. However, the remaining optimization problem is still non-convex due to the multiplication of optimization variables, i.e., $a_k$ and $\mathbf{\Theta}$ in the objective function and constraints. To overcome this, we use the AO approach and optimize the NOMA power coefficients as below:
\begin{subequations}
\begin{align} 
(\mathrm{P1.1}):  & \max_{ a_k} \hspace{0.5em} \log_2\left(1 + \frac{\alpha a_k P_\text{T} |\q h \mathbf{\Theta} \q f_k|^2}{\sigma^2}\right),\\
 & \text{s.t.}  \quad \eqref{opt1:2},~\eqref{opt1:6}.
 \end{align}
 \end{subequations} 
Upon replacing \eqref{opt1:6} by \eqref{opt1:2}, we reach to $a_k\leq (1+\gamma_{th})^{-1} \left[1-\frac{\gamma_{th}\sigma^2}{\alpha P_T |\q h \mathbf{\Theta} \q f_{\bar{k}}|}\right] 
$. For the objective to be maximized, $a_k$ must equal its upper bound:
\begin{align}\label{opt_ak}
    a_k= (1+\gamma_{th})^{-1} \left[1-\frac{\gamma_{th}\sigma^2}{\alpha P_T |\q h \mathbf{\Theta} \q f_{\bar{k}}|}\right],~\forall k\in \mathcal{K}.
\end{align}This indicates the allocation of power to the strong user. As a result, for the weak user, it is $1-a_k$. Next, the optimization problem over $\mathbf{\Theta}$ can be rewritten as follows:
\begin{subequations}
\begin{align}
 (\mathrm{P1.2}):  & \max_{ \q v} \hspace{0.5em} \log_2\left(1 + \frac{\alpha a_k P_\text{T} |\q v \q H \q f_k|^2}{\sigma^2}\right),\\
 & \text{s.t.}  \quad
      |\q v \q H \q f_k|^2\geq |\q v \q H \q f_{\bar{k}}|^2,~\text{for}\hspace{0.5em} \Phi(k)=1, \Phi(\bar{k}) = 0,\\
 &\quad\quad \frac{\alpha a_{\bar{k}} P_\text{T} |\q v \q H\q f_{\bar{k}}|^2}{ \Phi(k)\alpha a_{k} P_\text{T} |\q v \q H \q f_{\bar{k}}|^2+\sigma^2}\geq \gamma_{\text{th}} ,~\text{for}~ \Phi(k)=1,\\
 &\quad\quad \frac{L(1-\alpha)P_\text{T}|\q v \q H\q f_{k}|^2}{\sigma^2} \geq \gamma_{\text{th}} ,~\forall k\in \mathcal{K},\\
 &\quad\quad |\q v|=1 ,~\forall m\in\mathcal{M}, 
\end{align}
\end{subequations}
where $\q v = \left[e^{j\theta_1},\dots, e^{j\theta_m},\dots,e^{j\theta_M}\right]$  and $\q H = \textrm{diag}\left(\q h\right)$. Nevertheless, $(\mathrm{P1.2})$ is non-convex as it contains a quadratic form over $\q v$. To address it, via the SDR technique, we  relax the non-convex problem by defining a new variable as $\q V = \q v^H \q v$ that  satisfies $\textrm{Rank}(\q V) = 1$ and $\mathbf{V}\succeq\mathbf{0}$. Hence, the decoding order constraint at the users (\ref{opt1:1}) can be expressed as
\begin{align}\label{eq10}
      \!  \mathrm{Tr}(\!\q H \q f_k \q f_k^H \q H^H\q V\!)\!\geq\!   \mathrm{Tr}(\q H \q f_{\bar{k}}  \q f_{\bar{k}}^H \q H^H \q V),\hspace{0.2em}\text{for} \hspace{0.2em}\Phi(k)\!  =  \!1, \Phi(\bar{k})\!  =\!  0,
\end{align} 
Finally, by dropping the non-convex rank-one constraint, (P1.2) can be reformulated as
\begin{subequations}
\begin{align}
 (\mathrm{P1.3}):  & \max_{ \q V} \hspace{0.5em} \log_2\left( 1 + \frac{\alpha a_k P_\text{T} \mathrm{Tr}(\q H \q f_k  \q f_k^H \q H^H \q V)}{\sigma^2}\right),\\
 & \text{s.t.}  \quad  \frac{\alpha a_{\bar{k}} P_\text{T} \mathrm{Tr}(\q H \q f_{\bar{k}}  \q f_{\bar{k}}^H \q H^H \q V)}{ \Phi(k)\alpha a_{k} P_\text{T} \mathrm{Tr}(\q H \q f_{\bar{k}}  \q f_{\bar{k}}^H \q H^H \q V)+\sigma^2} \geq  \gamma_{\text{th}} ,\label{13b}\\
 &\quad \quad \frac{L(1-\alpha)P_\text{T}\mathrm{Tr}(\q H \q f_k  \q f_k^H \q H^H \q V)}{\sigma^2} \geq \gamma_{\text{th}} ,~\forall k,\label{13d}\\
 &\quad \quad \eqref{eq10}, \quad    \text{diag}(\mathbf{V})=\mathbf{1}_{M}, \quad  \mathbf{V}\succeq\mathbf{0}.
\end{align}
\end{subequations}However, $ (\mathrm{P1.3})$ usually results in a solution with a rank higher than one. To obtain a suboptimal solution, we define a penalty term for the rank-one constraint \cite{Shayan_zar1}. 
For the positive semidefinite matrix $\mathbf{Y} \in \mathbb{H}^{N\times N}$, the rank-one constraint can be expressed
as the difference of two convex functions, i.e.,
\begin{equation}\label{32}
\textrm{Rank}(\mathbf{Y})=1 \Longleftrightarrow ||\mathbf{Y}||_*-||\mathbf{Y}||_2 = 0,
\end{equation}

where $||\mathbf{Y}||_*=\sum_{j} \delta_j$, $ ||\mathbf{Y}||_2=\underset{j}{\text{max}}\{\delta_j\}$, and $\delta_{j}$ is the $j$-th singular value of $\mathbf{Y}$.
Consequently, we apply the penalty-based approach by integrating such a constraint into the objective function of $(\mathrm{P1.3})$, denoted by $F(\q V)$. Thus, we have the following optimization problem:
\begin{subequations}
	\begin{align}
 (\mathrm{P1.4}):  &  \max_{ \q V} \hspace{0.5em}  F(\q V)- \frac{1}{2\mu}(||\mathbf{V}||_*-||\mathbf{V}||_2),\\
 & \text{s.t.} \quad \eqref{eq10},~\eqref{13b},~\eqref{13d}, \label{31b}\\
	&\quad\quad \text{diag}(\mathbf{V})=\mathbf{1}_{M}, \quad \mathbf{V}\succeq\mathbf{0},\label{31c}
	\end{align}
\end{subequations}
where $\mu$ is a penalty factor for (\ref{32}). Specifically, for a sufficiently small value of $\mu$, solving $ (\mathrm{P1.4})$ yields a rank-one solution \cite{Shayan_zar1}. However, $ (\mathrm{P1.4})$ is still not a convex optimization problem yet due to the difference of concave functions  (D.C.) form of the objective function. To address this,  we define a lower bound for $\Delta=||\mathbf{V}||_2$ from  its first-order Taylor series expansion, which is given by
\begin{align}
\!\!{\small \Delta(\mathbf{V})\geq \Delta(\mathbf{V}^t)+\text{Tr}\bigg(\nabla_{\mathbf{V}}^H\Delta(\mathbf{V}^{t})(\mathbf{V}-\mathbf{V}^{t})\bigg)\triangleq\tilde{\Delta}(\mathbf{V})}.
\end{align}\begin{algorithm}[t]
 		\caption{Alternating Optimization (AO) Algorithm}
 		\begin{algorithmic}[1]\label{algorithm_AO}
 			\renewcommand{\algorithmicrequire}{\textbf{Input:}}
 			\renewcommand{\algorithmicensure}{\textbf{Output:}}
 			\REQUIRE Initialize the number of iterations $i$, acceptable tolerance, $ \epsilon  \ll 1$, random phases, $\mathbf{\mathbf{\Theta}}^{(i)}$, random NOMA power coefficients, $a_k^{(i)}$, and $R_k^{(i)} = 0$.\\
 			\STATE \textbf{repeat}\\
 	 	 \STATE \quad For given $\mathbf{\Theta}=\mathbf{\Theta}^{(i)}$ and $a_k^{(i)}$, calculate $\alpha_{\text{opt}}^{(i)}$   
             from    \eqref{alpha_opt}.
 			\STATE \quad  Solve \eqref{opt_ak} to obtain $a_k^{(i+1)}$. \\ 
 			\STATE \quad  Solve $(\mathrm{P1.4})$ to obtain  $\mathbf{V}^{(i+1)}$ using  Algorithm 1  in  \\ \quad \cite{Shayan_zar3}. \\ 
 			 \STATE \quad  Decompose $\mathbf{V}^{(i+1)}=\mathbf{v}^{(i+1)}(\mathbf{v}^{(i+1)})^H$ and update\\ \quad $\mathbf{\Theta}^{(i+1)}=\text{diag}(\mathbf{v}^{(i+1)})$.
 			\STATE \quad Set $i=i+1$;
 			\STATE  \textbf{until} $|R_k^{(i)}-R_k^{(i-1)}|<\epsilon$.
 		\end{algorithmic}
\end{algorithm}  The transformed problem  $(\mathrm{P1.4})$ is a standard semi-definite programming (SDP) that can be solved efficiently by using CVX \cite{CVX}.

\section{Simulation Results}\label{numerical}
This section presents numerical results to evaluate the performance of \textbf{Algorithm \ref{algorithm_AO}}. The IRS comprises a  two-dimensional uniform rectangular array of phase shifts. All users are randomly located in the $[2\colon\!20,1\colon\!2]$ meters (m). The AP location and the IRS location are considered as $(0,0)$ m and $(2,2)$ m, respectively. The Rician factor is set to $3$ dB, $L = 10$, $\gamma_{\text{th}} = 10$ dB \cite{Park2020}, $\sigma^2 = -110$ dBm, and $\mu=5\times 10^{-5}$ \cite{Shayan_zar1}. The average channel attenuation at a unit reference distance with $f = 915$ MHz  is  $\left({3\times 10^8}/{4\pi f}\right)^2 d^{-\xi}$, where $d$ is the distance between nodes and $\xi=2.1$ is the pathloss exponent \cite{Jia2020}. For comparison, three benchmark system designs are studied, namely, i) Benchmark $1$: \textbf{Algorithm \ref{algorithm_AO}} with random phase shifts, $\bf{\Theta}_{\text{rnd}}$;  ii) Benchmark $2$: \textbf{Algorithm \ref{algorithm_AO}} with OMA;  iii) Benchmark $3$: \textbf{Algorithm \ref{algorithm_AO}} with OMA and random phase shifts. These benchmarks allow us to discern the effect of not optimizing the phase shifts and not using NOMA.  The  OMA scheme is implemented as time division multiple access (TDMA) with equal transmission time to serve users. To maximize the SNR for each user,  the optimal phase control policy can be achieved by aligning the phase of the IRS to match with the phase of the cascaded channels, i.e., $\q h$ and $\q f_k$.
 %%%%%%%%%%%%%%%%%%%%%%%%%
\begin{figure}[t]
    \centering
   \includegraphics[width=3in]{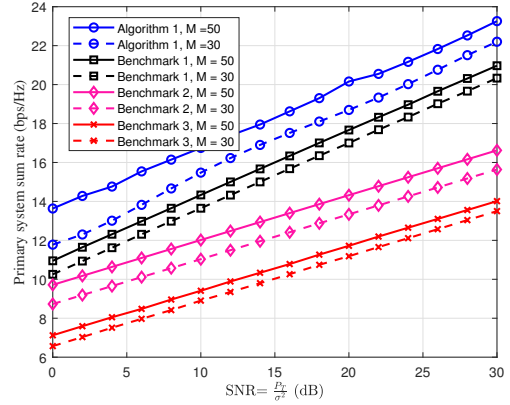}
    \caption{Primary system sum rate versus SNR.}   
    \label{Sum_rate_SNR}
\end{figure}
%%%%%%%%%%%%%%%%%%%%%%%%%

\begin{figure}[t]
    \centering
   \includegraphics[width=3in]{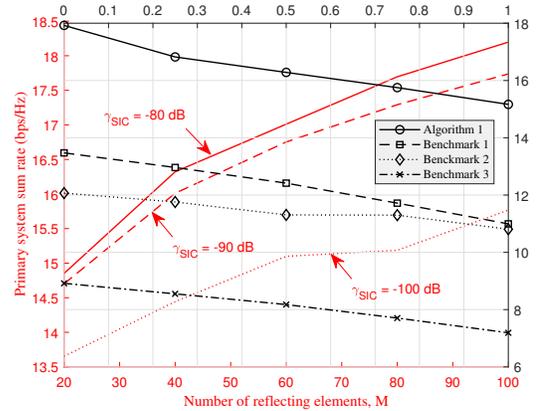}
    \caption{Primary  sum rate versus  $M$ with $P_\text{T} = 10$ dBm and CSI imperfection $\eta$ .}
    \label{reflecting_elements}
 
\end{figure}
%%%%%%%%%%%%%%%%%%%%%%%%%

 Figure \ref{Sum_rate_SNR} shows the impact of $\text{SNR} = \frac{P_T}{\sigma^2}$ (dB) on the sum rate of the primary system for two different numbers of phase shifts, $M$.  The figure shows that \textbf{Algorithm \ref{algorithm_AO}} outperforms other schemes. The impact of the optimal phase shifts is essential for performance. However, Benchmark $1$ has also better performance compared to the OMA transmission scheme, indicating the effectiveness of NOMA  even without optimized IRS phase shifts. Specifically, when $M = 30$, \textbf{Algorithm \ref{algorithm_AO}} improves the sum rate by $40\%$ and $68\%$ compared to Benchmark $2$ and Benchmark $3$, respectively. Furthermore, as the number of phase shifts increases, all the schemes achieve a higher sum rate. Indeed, the greater the number of phase shifts, the higher the number of multipath components, which improves the sum rate.

Figure \ref{reflecting_elements} illustrates the impact of imperfect CSI and imperfect SIC on \textbf{Algorithm \ref{algorithm_AO}}. The channel estimation model is given  as $\hat{\q h}=\q h+ \mathcal{\q e}$, where $\q h$ is the actual channel and $\mathcal{\q e}$ is the estimation error that is Gaussian distributed and zero mean, i.e., $\mathcal{\q e} \sim  \mathcal{N}(0,\,\sigma_{\mathcal{\q e}}^2)$. Error variance satisfies $\sigma_{\mathcal{\q e}}^2 \triangleq \eta |\q h|^2$, where $\eta$ controls  the level of CSI error. The right figure shows the primary system sum rate versus the $\eta$. As the CSI error increases, the sum rate of all schemes decreases. For instance, the performance loss is  $\%7.8$  with $\eta = 0.5$ compared to the ideal CSI case (i.e. $\eta = 0$). Now, let us consider the impact of imperfect SIC. The strong user rate is then $\log_2\left(1+ \frac{a_k \alpha P_T |\q h \mathbf{\Theta} \q f_k|^2}{\beta a_{\bar{k}} \alpha P_T |\q h \mathbf{\Theta} \q f_k|^2+\sigma^2}\right)$, where $\beta \in [0,1]$ denotes the SIC imperfection factor. For this simulation, we set $\beta =0.1$. On the other hand, to control this destructive factor, we replace the residual SIC term with a constant $\gamma_{SIC}$ in which it should satisfy the new constraint $\beta a_{\bar{k}} \alpha P_T |\q h \mathbf{\Theta} \q f_k|^2\leq \gamma_{SIC}$. As a result, the left figure shows the achieved sum rate versus $M$ for different residual SIC thresholds. The rate is sacrificed as $\gamma_{SIC}$ decreased because the phase shift needs to maximize the objective and satisfy the QoS while restricting the residual SIC.

\begin{figure}[t]
    \centering
   \includegraphics[width=3in]{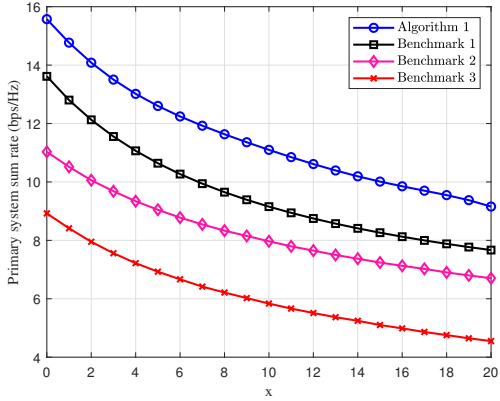}
    \caption{Primary  sum rate versus $x$ ($M=30$ and $P_T = 10$ dBm).}   
    \label{Sum_rate_distance}
\end{figure}

Figure \ref{Sum_rate_distance} shows the primary system sum rate versus $x$, where AP is located at $(-x,0).$  While keeping the location of the IRS and users fixed, we increase $x$. The resulting higher path loss decreases the primary sum rate. We observe that optimizing $\mathbf{\Theta}$ in \textbf{Algorithm \ref{algorithm_AO}} yields a better sum rate compared to the random and the OMA cases. The gap between \textbf{Algorithm \ref{algorithm_AO}}  with optimized and random $\mathbf{\Theta}$ is clear in both NOMA and OMA transmission. It highlights the spectrum efficiency advantage of  NOMA  with optimized phase shifts.

%%%%%%%%%%%%%%%%%%%%%%%%%
%%%%%%%%%%%%%%%%%%%%%%%%%
\section{Conclusion}  \label{conclusion}
This letter proposed and studied a two-user network that is served by an AP and IRS.   The  IRS serves dual functions as a conventional relay and backscatters its own data. The AP splits its transmit power between modulated and unmodulated signal parts to enable this process. The  IRS uses the latter to convey its data to the users. This setup creates  NOMA-based primary and exogenous secondary systems. We optimized the PS factor, IRS phase shifts, and NOMA power coefficients to maximize the rate of the strongest user in the primary while considering the decoding order at the users and satisfying QoS parameters for both weak user and IRS data of the primary and secondary systems, respectively. Our proposed algorithm achieves significant performance gains. Future extensions of this work include the multi-user case  ($>2$) and the multi-antenna AP case. Moreover, the energy efficiency can be optimized, yielding insights into the design of greener communication networks.  The proposed algorithm can be extended to consider the direct links between  the AP and the users. 
\bibliographystyle{ieeetr}
\bibliography{ref}

\end{document}